\documentclass[prl,twocolumn,showpacs]{revtex4}
\usepackage{graphicx,graphics,psfrag,amsmath,calc,amssymb}

\begin{document}
\title{Are Vortex Quasi-Crystals New Phases of Vortex Matter?}
\author{Wei Zhang}
\author{C. A. R. S\'{a} de Melo}
\affiliation{School of Physics, Georgia Institute of Technology,
Atlanta, Georgia 30332}

\date{\today}

\begin{abstract}

There seems to be a one to one correspondence between the phases of atomic
and molecular matter (AMOM) and vortex matter (VM) in superfluids
and superconductors.
Crystals, liquids and glasses have been experimentally observed
in both AMOM and VM. However, quasi-crystals also exist in AMOM, thus
a new phase of vortex matter is proposed here: the vortex quasi-crystal.
It is argued that vortex quasi-crystals are stabilized due to imposed
quasi-periodic potentials in large samples or due to boundary
and surface energy effects for samples of special shapes and sizes.
For finite size samples, it is proposed that a phase transition
between a vortex crystal and a vortex quasi-crystal occurs
as a function of magnetic field and temperature as the sample size
is reduced.

\end{abstract}
\pacs{74.25.Op, 74.25.-q, 74.25.Dw}

\maketitle

%
%

The general subject of vortex physics in superconductors is quite interesting
since there seems to be a large variety of possible equilibrium vortex
phases in superconductors~\cite{crabtree&nelson}.
The term ``vortex matter'' has been coined to emphasize
the complexity and diversity of vortex phases in superconductors
when compared to atomic and molecular matter.
One can think of a one to one correspondence between
phases in atomic and molecular matter (AMOM) and phases in vortex matter
(VM). A liquid in AMOM corresponds to a vortex liquid in VM; a crystalline
lattice in AMOM corresponds to a vortex lattice in VM; an amorphous or glassy
solid in AMOM corresponds to an amorphous or glassy vortex system in VM.
However, among all the possibilities discussed in
the vortex matter literature, a very interesting one is missing: the vortex
quasi-crystal. Quasi-crystals in AMOM were experimentally discovered
several years ago~\cite{experiments}, but there are no corresponding
experiments for vortex matter. Thus, the present paper is dedicated
to the proposal of vortex quasi-crystalline phases.
It will be argued that vortex quasi-crystals
maybe stabilized either by imposing quasi-periodic potentials or
by boundary effects and surface energies in finite systems.

The central question of this manuscript is:
are vortex quasi-crystals new phases of vortex matter? If so, what suggestions
can be given to experimentalists to help in the search for such phases?
It must be emphasized that
these questions will be addressed here on a preliminary basis,
and further detailed work will be necessary. In order to consider the
possibility of a vortex quasi-crystal one must seek under what conditions a
quasi-crystalline arrangement is at all possible.
A definite possibility is to argue that a
stable vortex quasi-crystal can arise for an imposed quasi-periodic potential
in a multilayered structure, like for instance in the case of
a superconductor that is grown on top of a flat quasi-crystal surface. Another
possibility is to create a quasi-periodic optical lattice with
laser beams, trap and cool atoms, and produce vortex
quasi-crystals in Bose or Fermi superfluids.
This experiment is natural since the production of vortex lattices in superfluid Bose or Fermi
ultra-cold atoms has become standard~\cite{cornell-99, ketterle-05}.
However, in this manuscript, we concentrate on the possibility of
stabilizing vortex quasi-crystals only due to boundary effects in
finite superconducting samples, where the sample size and shape
play an important role. This option is motivated by recent
experimental studies of the vortex structure in disk,
triangular, square, and star-shaped mesoscopic samples~\cite{geim-97,geim-00,chibotaru-01,dikin-03}.
Present technology should allow the preparation of
pentagonal (pentagon-cylinders) or decagonal (decagonal-cylinders)
samples as potential candidates to produce 5-fold vortex quasi-crystals.
If boundaries are important then in the thermodynamic limit of infinite volume
it is generally difficult to produce vortex quasi-crystalline order,
unless a quasi-periodic potential is imposed, as mentioned above.

This possibility is considered here under the following program.
In this manuscript only the case of an isotropic
type II superconductor in a magnetic field is considered. First,
the bulk free energy is calculated for a triangular, a square and a 5-fold
quasi-crystal array. The 5-fold quasi-crystal array is modeled by a Penrose
tiling of the plane. It is shown that the
Penrose tile array (vortex quasi-crystal) has a bulk free
energy which is just a few percent higher than the triangular array.
Second, instead of considering an infinite
(bulk) system, a pentagon cylinder sample is discussed. The pentagon cylinder
has a pentagonal cross-section (in the $xy$ plane) with side dimension
$\ell$, and with height $L$ along the ${\bf z}$ direction.
In this case, when the sample size gets smaller
the contribution of the boundaries (surface energy) to the total free energy
of the system becomes more important.
The surface energy is highly sensitive to the symmetry and to the surface
area of the boundaries. Taking into account the surface free energies,
it is shown that the Penrose tile (vortex quasi-crystal) has lower
free energy than the triangular lattice in certain regions of
the magnetic field versus temperature phase diagram. This is
suggestive that a ``first order phase transition'' may occur between the
triangular lattice and the Penrose tiling
(vortex quasi-crystal)~\cite{footnote2}.

For simplicity, the starting point of the analysis to follow is
the Ginzburg-Landau (GL) free energy density
\begin{equation}
\label{eqn:fed}
\Delta F_s = F_1
+ \frac{1}{2m} \left\vert \left( -i\hbar \nabla - \frac{2e {\bf A}}{c} \right)
\Psi ({\bf r}) \right\vert^2 + \frac{H^2}{8 \pi}
\end{equation}
for a bulk isotropic superconductor with no disorder,
where $\Delta F_s = F_s - F_n$ is the free energy density
difference between the superconducting state $(F_s)$ and the normal state $(F_n)$,
and $F_1 = \alpha \vert \Psi ({\bf r}) \vert^2 + \beta  \vert \Psi ({\bf r}) \vert^4/2$.
It is useful to introduce the dimensionless quantities
$f = \Psi \sqrt{-\beta/\alpha}$, $\mbox{\boldmath$\rho$} = {\bf r}/ \lambda (T)$,
${\cal A} = 2 \pi \xi (T) {\bf A}/ \Phi_0 $, and
${\cal H} = 2 \pi \xi (T) \lambda (T) H / \Phi_0$,
where $\lambda$ is the penetration depth, $\xi$ is the coherence length,
and $\Phi_0 = hc/2e$ is the unit flux.
Here, $f = f_0 \exp (i\phi)$, and ${\cal A } = {\cal A}_0 +  {\nabla \phi / \kappa}$ with
$\kappa = \lambda /\xi$ is the GL parameter.
Notice that when $f_0 = 1$ the system is fully superconducting,
and when $f_0 = 0$ the system is normal, thus $f_0 \le 1$ always.
Considering the minimization of free energy density with respect to
${\cal A}$ and $\Psi$ it is easy to arrive at equations for
the dimensionless functions ${\bf {\cal H}}$ and $f_0$.
The microscopic field is
\begin{equation}
\label{eqn:hsolution}
{\cal H}(x,y) = \kappa \left[ 1 + \frac{ H - H_{c_2}}{H_{c_2} } \right]
- \frac{g(x,y)}{ 2 \kappa},
\end{equation}
with ${\bf H}$ parallel to the ${\bf z}$ direction,
and the equation for $f_0$ can be written as
\begin{equation}
\label{eqn:glinear}
\nabla^2 (\log g) + 2\kappa^2 = 0,
\end{equation}
where $g = f_0^2$ is a positive definite function.
Notice that ${\cal H} = \kappa$ for $H = H_{c_2}$ and $f_0 = 0$ $(g = 0)$.
The most general solution of Eq.~(\ref{eqn:glinear}) has the form~\cite{saint-james-69}
\begin{equation}
\label{eqn:gsolution}
g(x,y) = \exp\left[-\kappa^2(x^2 + y^2)/2\right] \exp \left[ \gamma(x,y) \right],
\end{equation}
where $\gamma(x,y)$ satisfies Laplace's equation $\nabla^2 \gamma (x,y) = 0$.
This means that $\gamma (x,y)$ is a harmonic function
excluding the locations of vortices, and can be expressed
as the real part of an analytic function of $z = x + iy$.
This observation has very important consequences
for the microscopic field profile ${\cal H} (x, y)$ of Eq.~(\ref{eqn:hsolution}),
which depends strongly on the structure of $g(x,y)$.

The bulk Gibbs free energy density can be expressed as~\cite{abrikosov}
\begin{equation}
\label{eqn:gibbs}
G_s (H,T) = G_n (H,T) - \frac{1}{8 \pi} \frac{ (H_{c_2} - H)^2} {(2 \kappa^2 - 1) \beta},
\end{equation}
where the parameter $\beta = {\langle g^2 \rangle/ {\langle g \rangle}^2}$
is a geometrical factor independent of $\kappa$.
The notation $\langle \cdots \rangle$ indicates average over volume.
It is important to notice that $\beta \ge 1$ no matter what is the
form of $g(x,y)$ because of the Schwartz inequality. In addition, notice
that the Gibbs free energy above is a minimum, whenever $\beta$ reaches its minimum value.

For the purpose of calculating the parameter $\beta$ and the free energies
corresponding to different vortex configurations,
the analytical structure of $g(x,y)$ in the complex plane is used to
rewrite it as
\begin{equation}
\label{eqn:gzzbar}
g(z,{\bar z}) = \exp (-\kappa^2 z {\bar z}/2) \vert P (z) \vert,
\end{equation}
where $P (z) = {\cal N} \prod_{i = 1}^M (z - z_i)$.
Here each $z_i$ corresponds to a zero of $g$ in the complex plane,
and $M$ is the number of zeros. The zeros $z_i$ indicate the location
of vortices. From now on it is assumed that there is only one vortex
with flux $\Phi_0$ at each position $z_i$, i.e., each zero in non-degenerate.
In this case, $M$ corresponds to the number of vortices,
and thus the total flux threading the sample is $\Phi = M \Phi_0$.
The normalization coefficient ${\cal N}$ just guarantees that $g(z,{\bar z}) \le 1$.
Depending on the locations of the zeros of $g(z, \bar z)$,
it is possible to study several possibilities of periodic and quasi-periodic vortex arrangements.
In this study only vortex crystals corresponding to
triangular and square lattices and vortex quasi-crystals corresponding
to the 5-fold Penrose tiling of the plane are considered.
Both the square lattices and triangular lattices can be generated via
the tiling method, i.e., via the periodic arrangements of identical
square tiles or identical lozenges of internal angles ($60^\circ$ and $120^\circ$).
The Penrose lattice, however, requires quasi-periodic
arrangements of two types of tiles (lozenges), one with
internal angles $36^\circ$ and $144^\circ$ and the other with internal
angles $72^\circ$ and $108^\circ$. Using the representation in Eq. (\ref{eqn:gzzbar}),
the values of $\beta$ for the triangular, square and Penrose tiling are
respectively $\beta_3 = 1.16$, $\beta_4 = 1.18$ and $\beta_5 = 1.22$. This
immediately indicates that the triangular lattice has lower free energy
than the square lattice which has lower free energy
than the 5-fold vortex quasi-crystal (Penrose tiling), as we expected.
However, the fact that the free energy difference
between the triangular and 5-fold vortex quasi-crystal is only a few percent
suggests that appropriate boundaries can favor 5-fold symmetry as the sample size
gets smaller, as can appropriately imposed quasi-periodic potentials.

\begin{figure}[ht]
\begin{center}
\includegraphics[width=5.0cm]{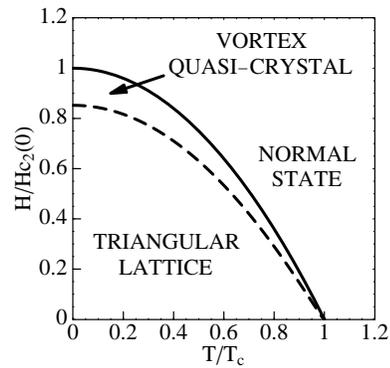}
\caption{$H - T$ phase diagram for a pentagonal cylinder cut out of
a regular cylinder of radius for $R_e = 10^{-6}$m.
The solid and dashed lines represent $H_{c_2}$ and $H_Q$, respectively,
and the superconductor is assumed to have $\kappa = 20$, $H_{c_2} (0) = 10$T.
}
\label{fig:1}
\end{center}
\end{figure}

In order to investigate how boundary effects can modify the total free
energy of the system, a sample in the shape of a pentagon cylinder of
side $\ell$ and height $L$ is considered.
Imposing that no currents flow through the sample boundaries
leads to the condition
$
{\bf \hat n \cdot} \left[
{ { - i {\bf \nabla}} - {2\pi {\bf A} / \Phi_0}}\right] \Psi(x,y) = 0
$
at all five side faces. The unit vector ${\bf \hat n}$ points along the
normal direction of each facet of the pentagon cylinder.
The boundary conditions can be translated
in terms of the harmonic function $\gamma (x,y)$ as
\begin{equation}
\label{eqn:bound-gamma}
n_x \frac{\partial \gamma}{\partial y} -
n_y \frac{\partial \gamma}{\partial x}
= \kappa (n_x y - n_y x),
\end{equation}
where $n_x$ and $n_y$ are the $x$ and $y$ components of the normal unit
vector ${\bf \hat n}$ at each one of the pentagon cylinder side faces.
The solution for this boundary value problem can be obtained
using a Schwarz-Christoffel conformal map of the pentagon into a semi-infinite plane
\begin{equation}
\label{eqn:sc-transf}
\frac{dz}{d w} = C (w - x_1)^{-2/5} (w^2 - x^{2}_2)^{-2/5}
(w^2 - x^{2}_3)^{-2/5},
\end{equation}
where the vertices of the pentagon located at $z_{i}$ are mapped into the
points $(x_1, \pm x_2, \pm x_3)$ on the real axis of the $w$-plane.
The full solution of this problem is complicated,
and requires heavy use of numerical methods~\cite{henrici}.
However, the free energy density can be estimated when the bulk solution
in Eq. (\ref{eqn:gzzbar}) is treated as a variational solution of
the boundary value problem determined by Eqs. (\ref{eqn:glinear}),
(\ref{eqn:bound-gamma}) and (\ref{eqn:sc-transf}).
The Gibbs free energy density difference $\Delta G = G_3 - G_5$
between the triangular and the 5-fold quasi-periodic Penrose structure
then becomes
\begin{equation}
\label{eqn:gibbs-dif}
\Delta G = - \frac{1}{8\pi} \frac{(H_{c_2} - H)^2}{(2\kappa^2 - 1) \beta^*}
+ \frac{H_c^2}{4 \pi} \epsilon(H),
\end{equation}
where $\beta^* = \beta_3 \beta_5/(\beta_5 - \beta_3)$,
$H_c$ is the thermodynamic critical field, and
$\epsilon= (\alpha_3 - \alpha_5)(2 R_e + \alpha_3 + \alpha_5)/2 R_{e}^2$,
with $\alpha_3 = 0.93 a_0$, $\alpha_5 = 0.90 a_0$, and
$a_0  = \sqrt{\Phi_0/H} $.
The effective length of the sample $R_{e} = \ell/\sqrt{4 - \tau^2}$,
where $\tau = 2 \cos(\pi/5)$ is the golden mean.
This expression for $\Delta G$ is valid only when $H \gg \Phi_0/R_e^2$.
The second term in $\Delta G$ takes into account the boundary mismatch energy,
and indicates that as the size of the pentagon cylinder gets smaller
it becomes more favorable to have a 5-fold quasi-crystal rather than a regular
triangular lattice. Notice, however, that when $R_e \to \infty$ the
triangular lattice has lower Gibbs free energy as it must, and no transition
to a 5-fold quasi-crystal occurs. Thus, this possible transition may occur
for finite sized samples only. From the condition that $\Delta G = 0$ we obtain
\begin{equation}
\label{eqn:hq}
H_Q = H_{c_2}
\left[ 1 - \kappa^* \sqrt{ \beta^* \epsilon(H_Q)} \right],
\end{equation}
where the transition to a quasi-crystal occurs.
Here, $\kappa^* = \sqrt{2 \kappa^2 - 1}/\kappa$.
The phase diagram for a superconductor with $\kappa = 20$,
$H_{c_2} (0) = 10$T and $R_e = 10^{-6}$m is shown~\cite{hc3} in Fig.~\ref{fig:1}
using $H_{c_2} (T) = H_{c_2} (0) \left[ 1 - (T/T_c)^2 \right]$.

\begin{figure}[ht]
\begin{center}
\hspace{-1cm}
\includegraphics[width=5.5cm]{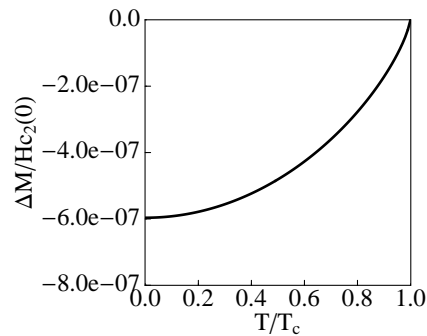}
\caption{The jump discontinuity $\Delta M = M_3 - M_5$ at
the critical field $H_Q$ for various reduced temperatures $T/T_c$,
using the same parameters of Fig. \ref{fig:1}.}
\label{fig:2}
\end{center}
\end{figure}

The jump of the magnetization $\Delta M (= M_3 - M_5)$
as a function of temperature at the critical field $H_Q$ can be calculated
from the Gibbs free energy leading to
\begin{equation}
\label{eqn:mag35}
\Delta M = \frac{H_{c_2}}{4\pi} \left [ -\frac{\gamma^*}{(2\kappa^2 -1)\beta^*}
+
\Delta M_s \right],
\end{equation}
where $\gamma^* = \kappa^* \sqrt{\beta^* \epsilon}$ and
$\Delta M_s = (\alpha_3 - \alpha_5)(R_e + \alpha_3 + \alpha_5)/4\kappa^2 R_e^2 (1-\gamma^*)$.
A plot of $\Delta M$ is illustrated in Fig. \ref{fig:2} for
the same parameters of Fig. \ref{fig:1}.
Using these parameters produces jump discontinuities
$\Delta M \approx - 0.060 $G at $T = 0$, and
$\Delta M \approx - 0.028 $G at $T = 0.8 T_c$.
(Measurements of $\Delta M$ may require the preparation of an
array of identical pentagonal cylinders to enhance the value.)
Notice that $\Delta M < 0$ indicates that the 5-fold vortex quasi-crystal
is denser than the triangular vortex lattice at $H_Q$,
being at best a few percent denser at $T = 0$.

In addition to magnetization measurements, it is also interesting
to perform calorimetric experiments. However, specific heat measurements
are very difficult because they require large samples.
Since sample size is important for the present discussion it
is not clear that such experiments can be successfully performed.
Nevertheless, the thermodynamic relationship between the magnetization and
entropy jumps is revealed in the Clapeyron equation
$\Delta S = S_3 - S_5 = - \Delta M dH_Q/dT$.
Since $d H_Q/ dT < 0$, and $\Delta M < 0$ implies that $\Delta S < 0$,
the entropy $S_3$ of the triangular vortex
lattice is less than the entropy $S_5$ of the 5-fold vortex quasi-crystal,
indicating that latent heat $L = T \Delta S$ is required to cause this phase transition.

Thermodynamic quantities can provide a good understanding of properties
averaged over the entire sample. However, the use of local probes is much
desired in order to reveal the change in structure from a triangular
vortex crystal to a 5-fold vortex quasi-crystal. Thus, neutron scattering,
Bitter decoration and scanning tunneling microscopy (STM)
experiments can help elucidate the structure of the vortex arrangement.

\begin{figure}[ht]
\begin{center}
\includegraphics[width=2.3cm]{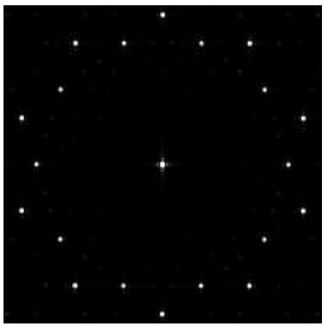}
\caption{First peaks of the square of the FT pattern for the 5-fold vortex
quasi-crystal (Penrose lattice). Notice the 10-fold symmetry revealed.}
\label{fivefold}
\end{center}
\end{figure}

In neutron diffraction experiments periodic or
quasi-periodic variations of ${\cal H} (x, y)$ will result in Bragg peaks.
The position of these peaks determine the characteristic length scale
of the vortex structure and its symmetry. The neutron scattering amplitude in
the Born approximation is
\begin{equation}
\label{eqn:scattering}
b ({\bf q}) = \frac{M_n} {2\pi \hbar^2}
\int \mu_n H ({\bf r}) \exp(i {\bf q \cdot r} ) d{\bf r},
\end{equation}
where $\mu_n = 1.91 e \hbar/M_n c$ is the neutron magnetic moment and the
$M_n$ is the neutron mass. The scattering amplitude $b ({\bf q})$ is directly
proportional to the Fourier transform $H ({\bf q})$ [${\cal H} (q_x, q_y)$]
of the microscopic field $H ({\bf r})$ [${\cal H} (x,y)$] of Eq. (\ref{eqn:hsolution}).
The neutron scattering cross section
$\sigma (q_x, q_y) = 4\pi^2 \vert b (q_x, q_y) \vert^2$
has sharp peaks at $ (q_x, q_y) = (0,0) $ (central peak)
and at $(q_x, q_y) = (\pm q_{xNm}, \pm q_{yNm}) $ (first Bragg peaks),
where $q_{xNm} = Q_N \cos (m\pi/N)$ and $q_{yNm} = Q_N \sin (m\pi/N)$,
with $ m = 0, 1, ..., N - 1$. For the triangular lattice
$N = 3$ the first Bragg peak occurs at
$|Q_3| = 2.31 \times \pi/d_3$, where $d_3$ is the lattice spacing.
For the 5-fold vortex quasi-crystal
(Penrose Lattice, $N = 5$) the first Bragg peak occurs at
$|Q_5|= 2.46 \times \pi/ d_5$, where $d_5$ is the side of a tile.
Since the sample size is important for the observation of a 5-fold
quasi-crystal, neutron scattering experiments may be difficult to
perform. Thus, Bitter decoration or STM may be better techniques. For
instance, STM scans at different fields and temperatures
in the vicinity of $H_Q (T)$
should reveal the real space locations of vortices,
which can be Fourier transformed (FT) to obtain a 6-fold pattern
for the triangular vortex lattice and a 10-fold pattern for
the 5-fold vortex quasi-crystal (Penrose lattice), see Fig.~\ref{fivefold}.

\vskip -2mm
Now that the phase diagram, thermodynamics, and the local signatures
of a vortex quasi-crystal have been discussed, it is important to say a word on
the stability of such structures. A stability analysis in the free
energy can be performed by moving the vortices away from their
equilibrium positions $z_i$ to $z_i + \delta z_i$.
The eigenvalues associated with these displacements indicates that
for $H > H_Q (T)$ the vortex quasi-crystal lattice is
stable. However, there is no full compatibility of the 5-fold
vortex lattice with the pentagon cylinder geometry, and thus the appearance
of disclinations and dislocations is possible. As a result there is an
additional possibility of the coexistence of a solid
(crystal or quasi-crystal) vortex phase in the center of the sample
and a liquid vortex phase closer to the edges.

In conclusion, we have shown that vortex quasi-crystals may be experimentally
observed in isotropic type II superconductors,
provided that the sample size and shape are properly chosen.
This result opens the possibility of a new phase of vortex matter:
the vortex quasi-crystal. This yet non-observed phase adds to the many phases
of vortex matter already seen experimentally, the vortex crystal,
the vortex liquid~\cite{liquid-1,liquid-2}
and the vortex glass~\cite{glass}.
By taking into account boundary effects, sample shape and size
or properly imposed quasi-periodic potentials, the present work suggests
that a vortex quasi-crystal phase may exist and compete with vortex liquid,
glass or crystal phases, as magnetic field, temperature and disorder are varied.

We would like to thank Wai Kwok for references, the Aspen Center for Physics
for their hospitality, and NSF (DMR-0304380) for financial support.


\begin{references}
\bibitem{footnote2} The term ``first order phase transition'' appear in
quotes since the sample size is finite.
\bibitem{hc3} The critical field $H_{c_3} (T)$ for surface superconductivity
is not shown in Fig.~\ref{fig:1}.
\end{references}
\end{document}